\documentclass{bmcart}

\usepackage{amsthm,amsmath}
\RequirePackage[authoryear]{natbib}
\RequirePackage{hyperref}
\usepackage[utf8]{inputenc} 

\usepackage{enumitem}
\usepackage{multirow}
\usepackage{amssymb}
\usepackage{relsize}
\usepackage{graphicx}

\startlocaldefs
\endlocaldefs

\begin{document}

\begin{frontmatter}

\begin{fmbox}
\dochead{Research}


\title{Uncovering Complex Overlapping Pattern of Communities in Large-scale Social Networks}


\author[
   addressref={aff1},                   
   corref={aff1},                       
   email={hwxu@link.cuhk.edu.hk}   
]{\inits{EHW}\fnm{Elvis H.\,W.} \snm{Xu}}
\author[
   addressref={aff1},
   email={pmhui@phy.cuhk.edu.hk}
]{\inits{PM}\fnm{Pak Ming} \snm{Hui}}

\setattribute{authorinfo}       {text} {Correspondence: hwxu@link.cuhk.edu.hk\\Department of Physics, The Chinese University of Hong Kong, Shatin, Hong Kong, China}


\address[id=aff1]{
  \orgname{Department of Physics, The Chinese University of Hong Kong}, 
  \street{Shatin},                     %
  \city{Hong Kong},                              
  \cny{China}                                    
}



\end{fmbox}


\begin{abstractbox}

\begin{abstract} 
The conventional notion of community that favors a high
ratio of internal edges to outbound edges becomes invalid when
each vertex participates in multiple communities. Such a
behavior is
commonplace in social networks.  The significant overlaps among
communities make most existing community detection algorithms
ineffective.  The lack of effective and efficient tools resulted
in very few empirical studies on large-scale detection and
analyses of overlapping community structure in real social
networks.  We developed recently a scalable and accurate method
called the Partial Community Merger Algorithm (PCMA) with linear
complexity and demonstrated its effectiveness by analyzing two
online social networks, Sina Weibo and Friendster, with $79.4$ and
$65.6$ million vertices, respectively. Here, we report in-depth
analyses of the $2.9$ million communities detected by PCMA to uncover
their complex overlapping structure.
Each community usually overlaps with a significant number of other
communities and has far more outbound edges than internal edges.
Yet, the communities remain well separated from each other.  Most
vertices in a community are multi-membership vertices, and they
can be at the core or the peripheral.  Almost half of the entire
network can be accounted for by an extremely dense network of
communities, with the communities being the vertices and the
overlaps being the edges.
The empirical findings ask for rethinking the notion of community,
especially the boundary of a community. Realizing that it is how
the edges are organized that matters, the $f$-core is suggested as
a suitable concept for overlapping community in social networks.
The results shed new light on the understanding of overlapping
community.
\end{abstract}



\end{abstractbox}
%

\end{frontmatter}




\section*{Introduction}

A community in networks is conceived commonly as a group of
vertices connected closely with each other but only loosely to the
rest of the network.  Such communities are widespread in many
systems and their detection has attracted much attention in the
past two decades~\citep{Fortunato:2010iw}. This vague notion of
communities is subjected to many possible interpretations. The most common one
is based on the ratio of the numbers of internal edges to outbound
edges, which go out of the community.  The more the internal edges
to outbound edges, the more definite is the community. For
example, the widely used methods based on strong/weak
community~\citep{Radicchi:2004br}, LS-set~\citep{Luccio:1969hu},
conductivity and network community profile~\citep{Leskovec:2009fy,
Jeub:2015hs}, and fitness functions~\citep{Baumes:2005tf,
Lancichinetti:2009dy, Goldberg:2010ij} favor a higher internal
edges to outbound edges ratio.  The idea works well for disjoint
communities, but it has also been adopted by algorithms for
detecting overlapping communities~\citep{Xie:2013ku}.  Nonetheless,
the number of members, mostly at the periphery, belonging to
multiple communities is still expected to be small so that an
``overlapping community" remains well separated from its
surrounding.
However, the structure of overlapping communities
in real social networks may be
far more complex.  It is commonplace that every individual has
multiple social circles.  It implies that all
parts of a social community, peripheral and core, may be
overlapping with a significant number of other communities and
there can be far more outbound edges than internal edges.
The existence of these significantly overlapped communities,
as will be shown in the present work, asks
for a deeper understanding of what an overlapping community really
is, where their boundaries are, and how to detect them.

Analyzing big data sets of real social networks is vital in
network science.  An immediate problem is that most existing
methods are incapable of detecting significantly overlapped groups
of vertices, because these groups have too many outbound edges to
be identified as well separated communities. The recently proposed
methods of OSLOM~\citep{Lancichinetti:2011gn} and
BIGCLAM~\citep{Yang:2013ko} are useful to some extent in small
synthetic networks, but they become inefficient for large-scale
networks which readily have the size of millions to billions of
vertices.  Sampling small subnetworks~\citep{Maiya:2010fx} would
not work either due to the small-world
effect~\citep{Watts:1998bj}, e.g. the average distance between any
two individuals on Facebook is only
$4.74$~\citep{Backstrom:2011vq, Ugander:2011ui}, while
in a social group for which the size is
small compared to the whole network, a member may usually need one
or two hops to be connected to all the other members.  A
community may be considered as localized, but it
is also widespread in
the network. Sampling small subnetworks would preserve particular
communities but decompose many others, making it inappropriate for
studying the overlaps among communities. Some newly proposed
algorithms~\citep{Lyu:2016ia, Sun:2017bw, Epasto:2017jw} achieved
linear-time complexity, but their validity and accuracy in
detecting significantly overlapped communities requires further
benchmarking and cross-checking.  The lack of effective and
efficient algorithms resulted in very few studies on detecting and
analyzing overlapping community structure in large-scale social
networks.  An empirical study was carried out on
Facebook~\citep{Ferrara:2012ic}, but only methods for detecting
disjoint communities were used. A recent study on Friendster found
that about $30\%$ of the vertices belonged to
multiple communities~\citep{Epasto:2017jw}. Jebabli et al analyzed
community structure in a sampled YouTube network of $1.1$ million
vertices and evaluated a number of overlapping community detection
algorithms~\citep{Jebabli:2015cq, Jebabli:2018kv}. Yang and
Leskovec analyzed metadata groups of some real networks and found
that overlaps occur more often at the cores of
communities~\citep{Yang:2015jw, Yang:2014fc}.  This is contrary to
the traditional notion that overlapping members are mostly at the
periphery.  Recent studies also revealed that metadata groups may
not give the ground-truth of structural
communities~\citep{Hric:2014jt, Peel:2017gp}.

The present authors developed recently a scalable partial
community merger algorithm (PCMA) which adopts $f$-core as the
notion of community that a member of a community should know at
least a fraction $f$ of the other members~\citep{Xu:2017gz,
Xu:2016ab}. The concept of $f$-core imposes no constraints or
implications on the fraction of overlapping vertices in a
community or the number of communities a vertex may belong to.
The method is a bottom-up approach by properly
reassembling partial information of communities found in ego
networks of the vertices to reconstruct the complete communities.
It consists of three steps:
\begin{enumerate}
\item Find communities in the ego
network of each of the vertices. These communities
are referred to as partial communities as each of them is only
part of the corresponding complete community.
\item Merge partial
communities that are parts of the same community to reconstruct
complete communities.
\item Clean up the noise
accumulated in the merged communities to sift out the real
communities.
\end{enumerate}
This approach is intuitive and easy to conceive.  There are a
number of similar  algorithms such as DEMON~\citep{Coscia:2012ip}
and EgoClustering~ \citep{Rees:2012jb}. The reason that PCMA
achieves a far better accuracy is a novel similarity measure of
communities that suppresses the amount of noise accumulated during
the merging process. The
present authors tested PCMA against the LFR
benchmark~\citep{Lancichinetti:2008ge} and a new benchmark
designed for significantly overlapping communities, and
established the accuracy and effectiveness of PCMA in detecting
communities with significant overlaps, as well as slightly
overlapping and disjoint ones. The linear complexity of PCMA
enabled the analysis of two huge online social networks with
$79.4$ and $65.6$ million vertices - Sina Weibo and Friendster
(see Table~\ref{tab:datasets}) - {\em without} sampling small
subnetworks. The $\sim 2.9$ million
communities detected by PCMA were verified to be
non-duplicating and have relatively high values of internal edge
density.  A surprising finding is that more than $99\%$ of them
have more outbound edges than internal edges, and the outbound
edges often outnumber the internal edges by many times.  The
communities overlap significantly, while still keeping relatively
clear boundaries. These communities are strong empirical evidence
against the traditional notion of an overlapping community. While
we focused on developing the PCMA algorithm in
Ref.~\citep{Xu:2017gz}, we uncover the complex overlapping pattern
of these communities in
the present work by examining the data in detail and explain why
the communities can still remain well separated from each other.
After introducing the four main characteristics of the overlapping
pattern, we give a macroscopic picture of the social network
structure by grouping edges of the entire network into five
types.  The concept
and possible better definitions of an overlapping community are
discussed. Additional information on the data sets and the
detection of communities is given in the appendix.

\begin{table}[t!]
  \caption{Information on the two huge social networks analyzed}
  \label{tab:datasets}
  \begin{tabular*}{\linewidth}{@{\hspace{\tabcolsep}\extracolsep{\fill}} lccccc}
    \hline
    Dataset  &  $n$  &  $m$ & $\left<k\right>$ & $C_{\mathrm{WS}}$ & $c$ \\
    \hline
    Sina Weibo  &  79.4M  &  1046M & 26.4 & 0.155 &  1.3M \\
    Friendster  &  65.6M & 1806M & 55.1 & 0.205 & 1.6M \\
    \hline
    \end{tabular*}

    \vspace{3pt}
    \raggedright
    M represents a million. $n$ and $m$ are the number of vertices and edges.
     $\left<k\right>$ is the average vertex degree.
     $C_{\mathrm{WS}}$ is the average local clustering coefficient.
     $c$ is the number of communities detected by PCMA.
     More detailed information is given in the appendix.
\end{table}

\section*{Characteristics of overlapping pattern}

In this section, we discuss in detail the four main
characteristics of the overlapping pattern of the $2.9$ million communities
detected by PCMA.

\begin{figure}[htbp]
   \centering
   \includegraphics[width=0.8\linewidth]{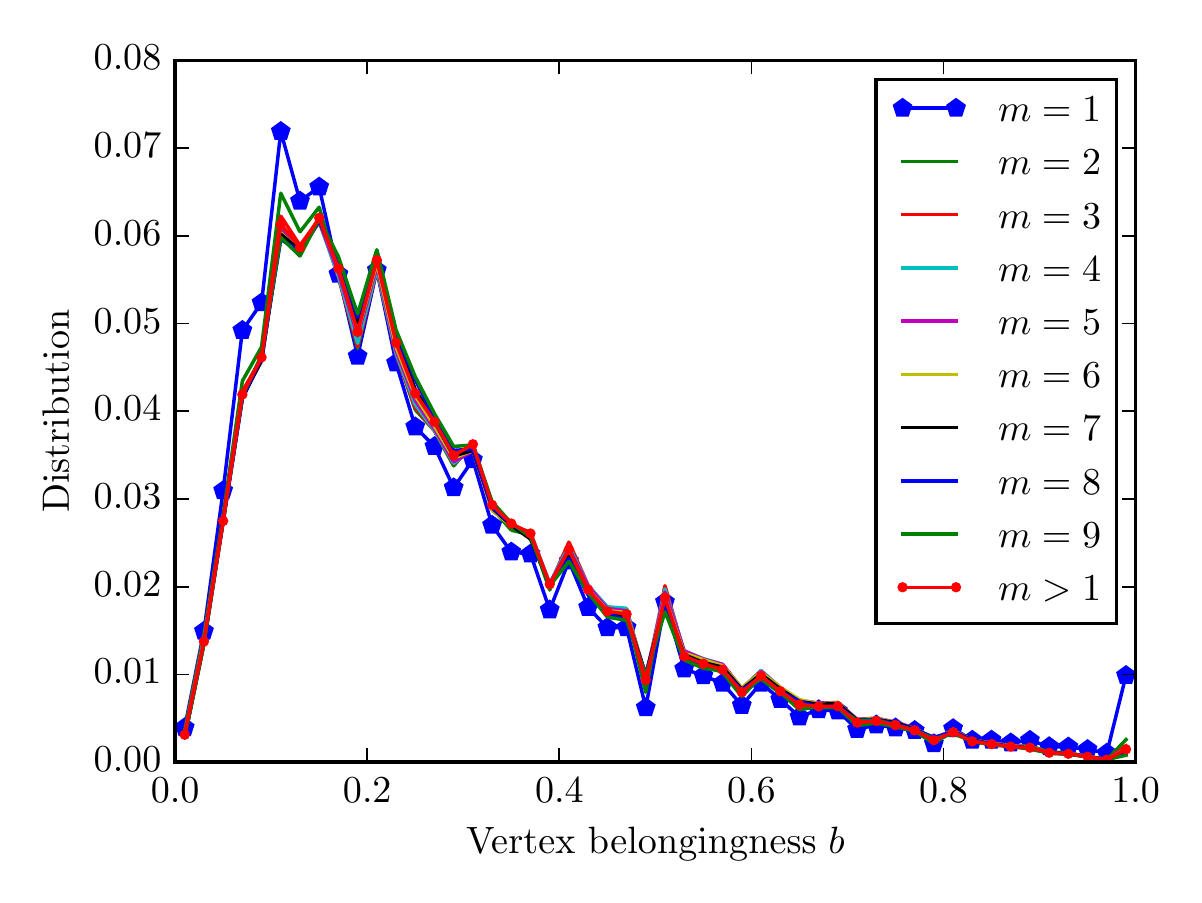}
   \caption{Belongingness distribution of vertices with
    different number of memberships in Sina Weibo network.
    The bin width of the x-axis is $0.02$.
    A vertex with $m$ memberships has $m$ independent values of
    belongingness $b$, each for a community that the vertex belongs to.
    The distributions are almost the
    same, i.e. uncorrelated to $m$, for different values of $m$,
    with $m=1$ shifted slightly to the left.
    The noisy peaks are due to the fact that belongingness $b$ as defined
    in Eq.~\eqref{eq:belongingness} is discrete, especially when $n_{\mbox{\tiny C}}$ is small.
    The Friendster network (not shown) shows the same pattern.}
   \label{fig:belongingness_membership}
\end{figure}

\textbf{\textit{Characteristic 1.}}
Multi-membership vertices or overlapping vertices account for the majority of the community, and they are everywhere.
These vertices were often thought to be
peripheral members. A recent
study on metadata groups~\citep{Yang:2014fc} found that these vertices are
more likely core
members. Our analysis on the two large-scale social networks
reveals that the overlapping vertices can be {\em anywhere}, i.e.,
core and periphery, in the community. In general, a vertex $v$ may
belong to $m_v$ communities.
The vertices can then be sorted by their values of $m_{v} = m$ for
$m \geqslant 1$.  The belongingness $b_{v,{\mbox{\tiny C}}}$ of a vertex $v$ to a
community $C$ can be defined as
\begin{equation}
b_{v,{\mbox{\tiny C}}}= \frac{k_{v,{\mbox{\tiny C}}}^{\mathrm{int}}}{n_{\mbox{\tiny C}}-1}
\label{eq:belongingness}
\end{equation}
where $n_{\mbox{\tiny C}}$ is the community size and $k_{v,{\mbox{\tiny C}}}^{\mathrm{int}}$ is
the number of other members in $C$ that are connected to $v$. A
high (low) value of $b_{v,{\mbox{\tiny C}}}$ means that $v$ is closer to the core
(periphery) of $C$.  If overlaps occur more often at the periphery
(core), we would expect multi-membership vertices with $m>1$ to
have a lower (higher) belongingness $b$ than those with $m=1$.
Fig.~\ref{fig:belongingness_membership} shows that the
belongingness distributions for vertices with different values of
$m$ are almost identical, with an insignificant tendency of
multi-membership vertices having a slightly higher belongingness.
The results imply that $m_{v}$ is basically uncorrelated with
$b_{v,{\mbox{\tiny C}}}$, and multi-membership vertices exist everywhere in a
community with no preference towards the core or the periphery as
compared with non-overlapping vertices.

\begin{figure}[htbp]
   \centering
   \includegraphics[width=0.8\linewidth]{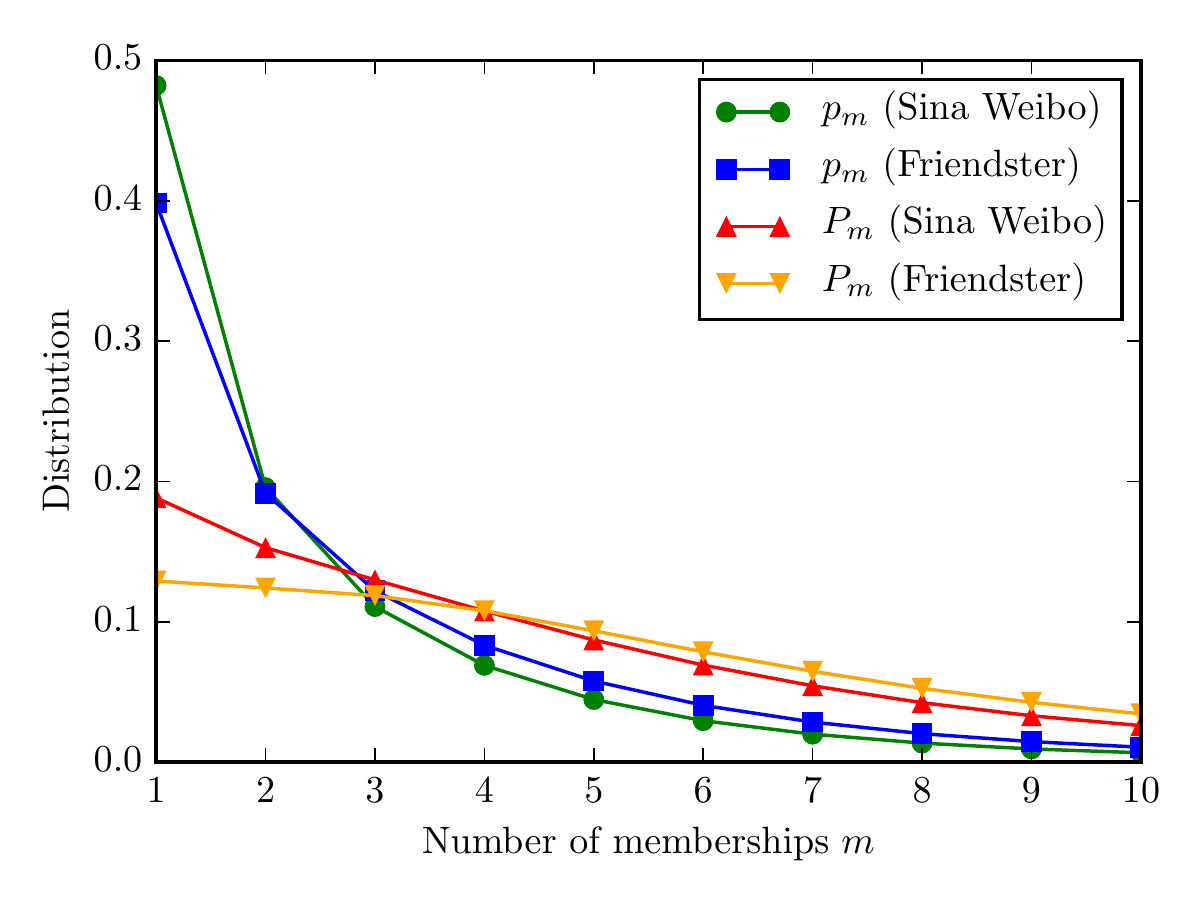}
   \caption{Membership distribution $p_m$ of vertices with $m \geqslant 1$ and
   the empirical probability $P_m$ that \textit{a member of a community} has $m$ memberships.
    $P_m$ is right skewed because a vertex with $m$ memberships is counted $m$ times (since it appears in $m$ communities) compared to $p_m$.
   }
   \label{fig:membership_dist}
\end{figure}

Fig.~\ref{fig:membership_dist} shows the distribution $p_{m}$ of the
number of memberships among the vertices with $m \geqslant 1$.
About $50\%$ of the
vertices, i.e., those with $m>1$, have multiple memberships in
Sina Weibo.  For Friendster, the proportion is $\sim 60\%$, which
is about twice of that reported in Ref.~\citep{Epasto:2017jw}.  A
related quantity is
\begin{equation}
P_m = \frac{p_m \cdot m}{\left<m\right>} \;,
\end{equation}
which gives the empirical probability that \textit{a member of a community} has $m$
memberships.  Here, $\left<m\right>=\sum_{m=1}^{\infty} p_m \cdot
m$ is the mean value of $m$.
Note that $P_m$ and $p_m$ are related but different.
$P_m$ is the expected membership distribution of the members within a
community,
and $p_m$ describes the distribution in $m$ of all
vertices with $m \geqslant 1$.
Referring to $P_{m}$ in
Fig.~\ref{fig:membership_dist}, $P_{m=1}=18.8\%$ and $12.9\%$ for
Sina Weibo and Friendster, respectively, implying that on average
more than $80\%$ of the members in a community are
multi-membership vertices.  This is in sharp contrast to the
preconceived idea that only a small fraction of members in a
community belong also to other communities.
The results reveal that most members of a community have multiple
memberships and they are everywhere in the community.

\begin{figure}[htbp]
   \centering
   \includegraphics[width=0.8\linewidth]{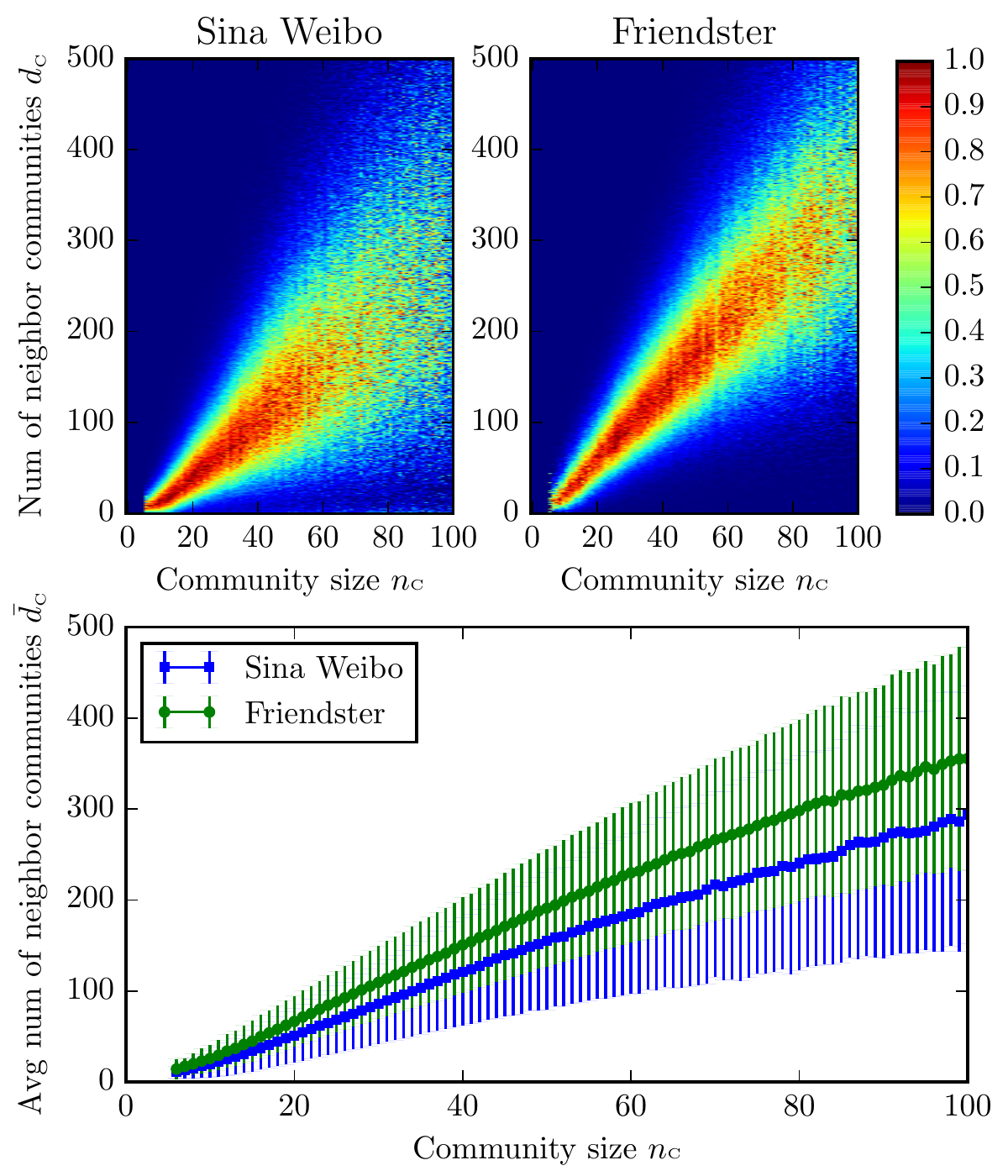}
   \caption{The relationship between a community's size
   and the number of its neighbor communities. Data are shown for community
   sizes from $6$ to $100$. The values in each vertical cut of the histograms
   are rescaled by mapping the highest value to unity. The bottom figure
   shows the averaged values and their standard deviations.}
   \label{fig:dist_m_c}
\end{figure}

\textbf{\textit{Characteristic 2.}} The multi-membership
vertices lead to a community overlapping with
many other communities.  We refer to them as \text{neighbor
communities}.  Fig.~\ref{fig:dist_m_c} shows the relationship
between the number of neighbor communities ${d_{\mbox{\tiny C}}}$ and the
size $n_{\mbox{\tiny C}}$ of a community in the two social networks.  To extract
information, the expected number of neighbor communities for a
community of size $n_{\mbox{\tiny C}}$ is roughly
\begin{equation}
  \bar d_{\mbox{\tiny C}} \left(n_{\mbox{\tiny C}}\right) \approx \left( \left<m\right>_{\mbox{\tiny C}} - 1 \right)
 \cdot n_{\mbox{\tiny C}}  \cdot r_{\mathrm{nd}}
\end{equation}
where
\begin{equation}
  \left<m\right>_{\mbox{\tiny C}} = \sum_{m=1}^{\infty} P_m \cdot m = \frac{\left<m^2\right>}{\left<m\right>}
\end{equation}
is the expected number of memberships of a member in the
community.  Although each member connects the community to
$\left<m\right>_{\mbox{\tiny C}} -1$ other communities of which it is also a
member, $\left( \left<m\right>_{\mbox{\tiny C}} - 1 \right) \cdot n_{\mbox{\tiny C}}$
overestimates the number of neighbor communities due to duplication, i.e.,
some members in the community have common neighbor communities. A
factor $r_{\mathrm{nd}}$ is introduced to represent the
non-duplicate rate.  Consider the simple case of a size-$n_{\mbox{\tiny C}}$
community with $x$ members all in only one neighbor community. In
this case, $\left( \left<m\right>_{\mbox{\tiny C}} - 1 \right) \cdot n_{\mbox{\tiny C}} = x$ while
$\bar d_{\mbox{\tiny C}} = 1$, implying $r_{\mathrm{nd}} = 1/x$.  Thus, the value
of $r_{\mathrm{nd}}$ also indicates the extent of overlap between
two communities.
For overlaps of just $2$ or $3$ vertices, $r_{\mathrm{nd}}$ drops
below $50\%$.
The analysis in Fig.~\ref{fig:dist_m_c} confirms that $\bar d_{\mbox{\tiny C}}
\sim n_{\mbox{\tiny C}}$, but with a slope gradually decreasing with increasing
$n_{\mbox{\tiny C}}$.  Thus, $r_{\mathrm{nd}}$ is negatively correlated with $n_{\mbox{\tiny C}}$.
The slopes are around $3\sim4$, which are about $30\%$ smaller
than the values $4.36$ and $5.25$ calculated by
$(\left<m\right>_{\mbox{\tiny C}}-1)$ from empirical data of Sina Weibo and
Friendster, respectively.  Note that these slopes are very large,
e.g. a community of size as small as $30$ could overlap with
$\sim100$ other communities concurrently. The resulting
non-duplicate rates $r_{\mathrm{nd}}$ are above $70\%$, strongly
indicating that most overlaps concern just one vertex.

\begin{figure}[htbp]
   \centering
   \includegraphics[width=0.8\linewidth]{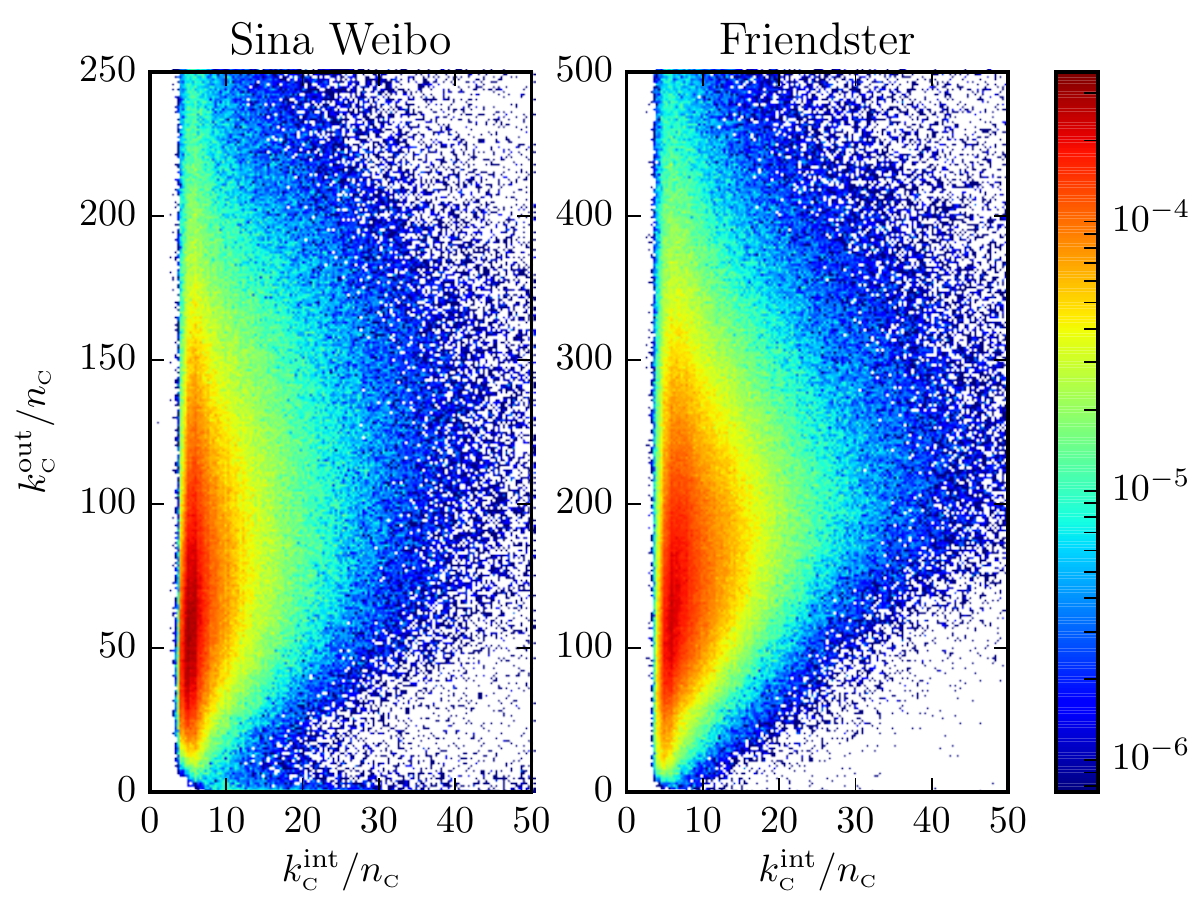}
   \caption{Histogram of communities grouped by the
   average internal and outbound edges per member. The numbers of bins
   in the $x$ and $y$ axes are $200$ and $400$, respectively.
   The counting in each bin is normalized by dividing the count by the total number of communities and the bin area. The bin area of the right panel is two times the left panel. To make the normalized values comparable to those in Figure~\ref{fig:hist_c_ex}, we set the bin area of the left panel to $1$.
    The normalized counting
   in each bin is given by the color, as defined by the color bar.
   More than $99\%$ of the detected communities have
   $k_{\mbox{\tiny C}}^{\mathrm{out}} > k_{\mbox{\tiny C}}^{\mathrm{int}}$, and  $k_{\mbox{\tiny C}}^{\mathrm{out}}$
   is usually much greater than $k_{\mbox{\tiny C}}^{\mathrm{int}}$.}
   \label{fig:ceio}
\end{figure}

\textbf{\textit{Characteristic 3.}} In contrast to
the generally believed notion that a community should have more
internal edges than outbound edges, we found that more than $99\%$
of the $2.9$ million communities have more outbound edges than
internal edges. For each community identified by PCMA, we
evaluated the total number of internal edges $k_{\mbox{\tiny
C}}^{\mathrm{int}}$ and outbound edges $k_{\mbox{\tiny
C}}^{\mathrm{out}}$:
\begin{equation}
 k_{\mbox{\tiny C}}^{\mathrm{int}} = \sum_{v \in C} k_{v,{\mbox{\tiny C}}}^{\mathrm{int}},\quad
 k_{\mbox{\tiny C}}^{\mathrm{out}} = \sum_{v \in C} k_{v,{\mbox{\tiny C}}}^{\mathrm{out}}
\end{equation}
where $k_{v,{\mbox{\tiny C}}}^{\mathrm{int}}$ ($k_{v,{\mbox{\tiny C}}}^{\mathrm{out}}$) denotes
the number of a vertex $v$'s edges that go inside (outside) the
community $C$.  The summations are over all $n_{\mbox{\tiny C}}$ vertices in
the community.  Note that each internal edge is counted twice as
both ends are within the community and each outbound edge is
included only once. Figure~\ref{fig:ceio} shows that the number of
outbound edges of a community is not only greater, but often many
times greater than the number of internal edges.  More than $99\%$
of the $2.9$ million communities have more outbound edges than
internal edges, in contrast to the traditional notion.

\begin{figure}[htbp]
   \centering
   \includegraphics[width=0.55\linewidth, trim={0 0.42cm 0 0.42cm}, clip]{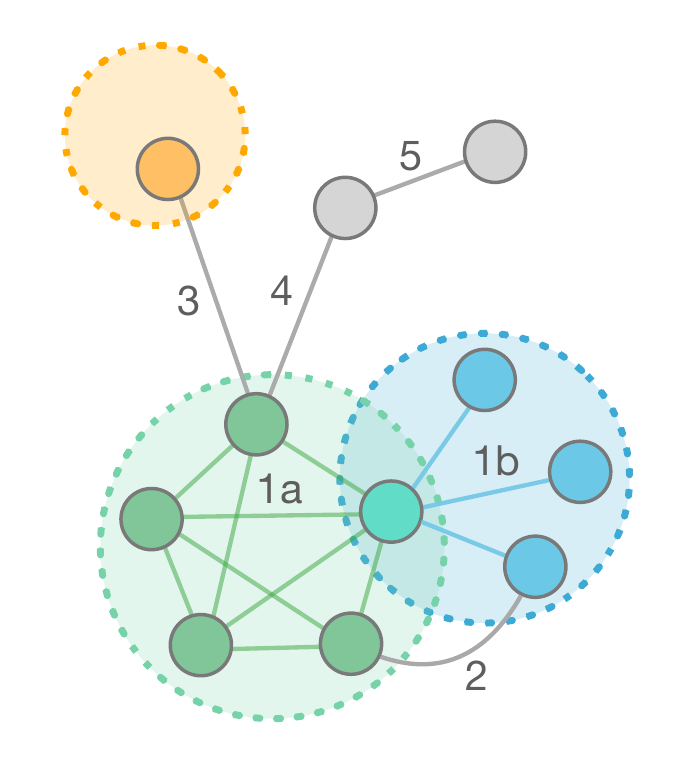}
   \caption{Edges can be classified into five {\em types}:
   (1) intra-community edges;
   (2) inter-community edges between two overlapped communities;
   (3) inter-community edges between two communities that do not overlap;
   (4) edges between vertices with membership $m>0$ and isolated vertices ($m=0$);
   (5) edges between isolated vertices.  Focusing on the outbound edges
   of the green community with $5$ members (circled), the edges 1b, 2, and 3+4
   correspond to {\bf categories} $E1$, $E2$, and $E3$
   outbound edges of the green community, respectively, as defined in the text.
   Different from the five types which classify edges of the whole network, the three categories $E1$, $E2$, and $E3$ are used to classify the outbound edges of a community and thus are introduced from the viewpoint of a particular community. There are some overlaps between the two types of classification: (1) An $E1$ outbound edge is by definition a Type $1$ edge; (2) A Type $2$ (Type $3$) edge is also an $E2$ ($E3$) outbound edge of the two corresponding communities that the edge connects. However, the reverse relationship is not always true. These types and categories are not interchangeable.
   }
   \label{fig:edge_classification}
\end{figure}

\begin{figure}[htbp]
   \centering
   \includegraphics[width=0.8\linewidth]{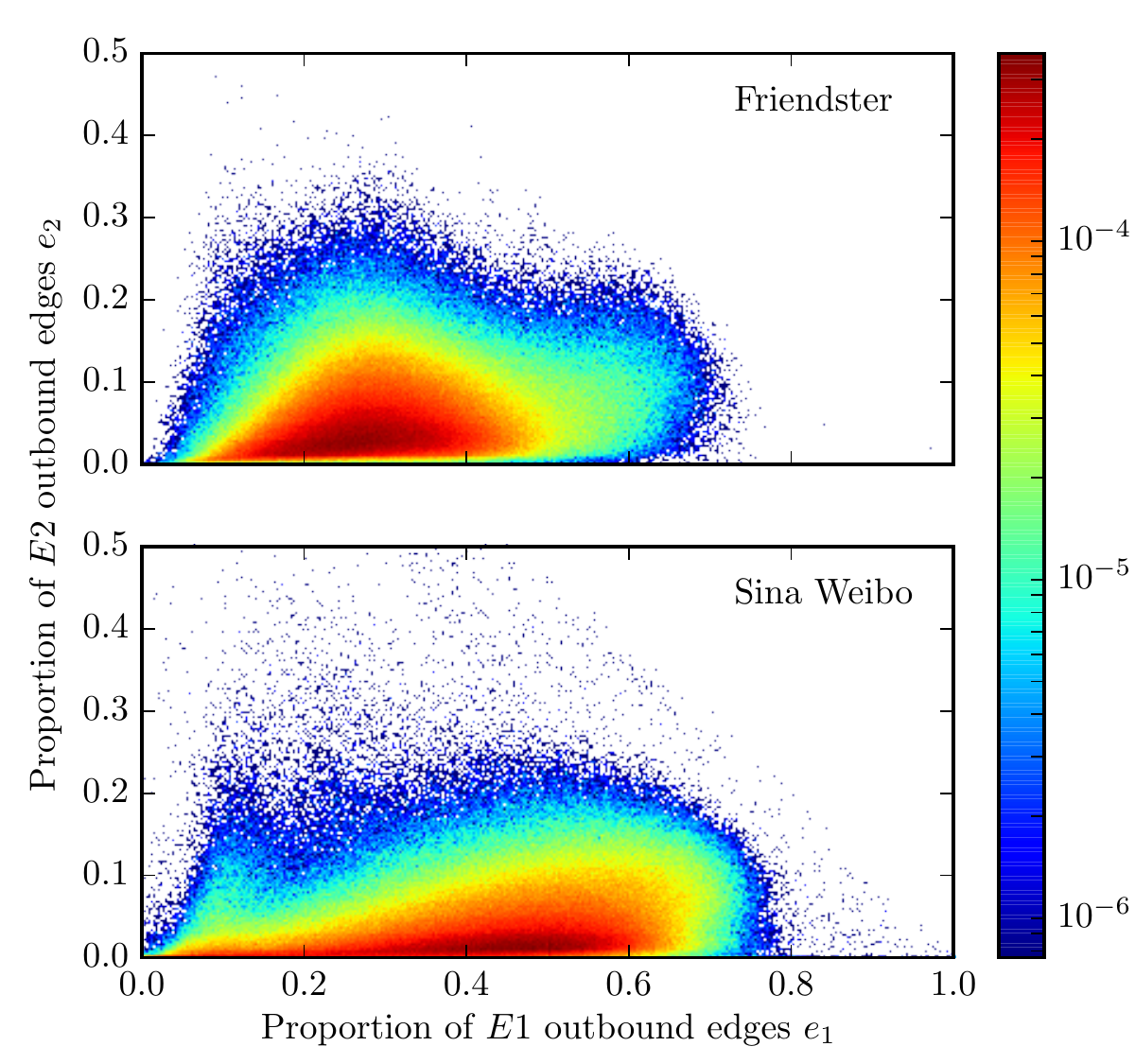}
   \caption{Histogram of communities grouped by the
    proportions of $E1$ and $E2$ outbound edges. The numbers of bins
    in the $x$ and $y$ axes are $400$ and $200$, respectively.  The counting
    in each bin is normalized by dividing the number by the total number of communities in each panel.}
   \label{fig:hist_c_ex}
\end{figure}

To investigate into the network structure, we focused on the
outbound edges and classified them into $3$ categories (see
Fig.~\ref{fig:edge_classification}) as
\begin{description}[noitemsep]
\item [$E1$:] outbound edges from a member to a neighbor community
to which the member also belongs; \item [$E2$:] outbound edges
from a member to a neighbor community that the member does not
belong to; \item [$E3$:] outbound edges not to a neighbor
community.
\end{description}
Their proportions $e_1$, $e_2$, $e_3$, with $e_1+e_2+e_3=1$, are
calculated for each community.  Fig.~\ref{fig:hist_c_ex} shows the
histograms.  Typically, the edges to a neighbor community are
usually through the common member(s) of the two communities as
$e_1$ is much greater than $e_2$.  In addition, a significant
proportion of outbound edges go to neighbor communities.  In Sina
Weibo, most communities (red region) have $e_1+e_2 \approx 0.5$.
It means that $\sim 50\%$ outbound edges are due to the
vertices' multi-membership and communities are densely connected
to their neighbor communities.  Note that if a community's
outbound edges were randomly connected to vertices in the network,
most edges would be of category $E3$.

\begin{figure}[htbp]
   \centering
   \includegraphics[width=0.8\linewidth]{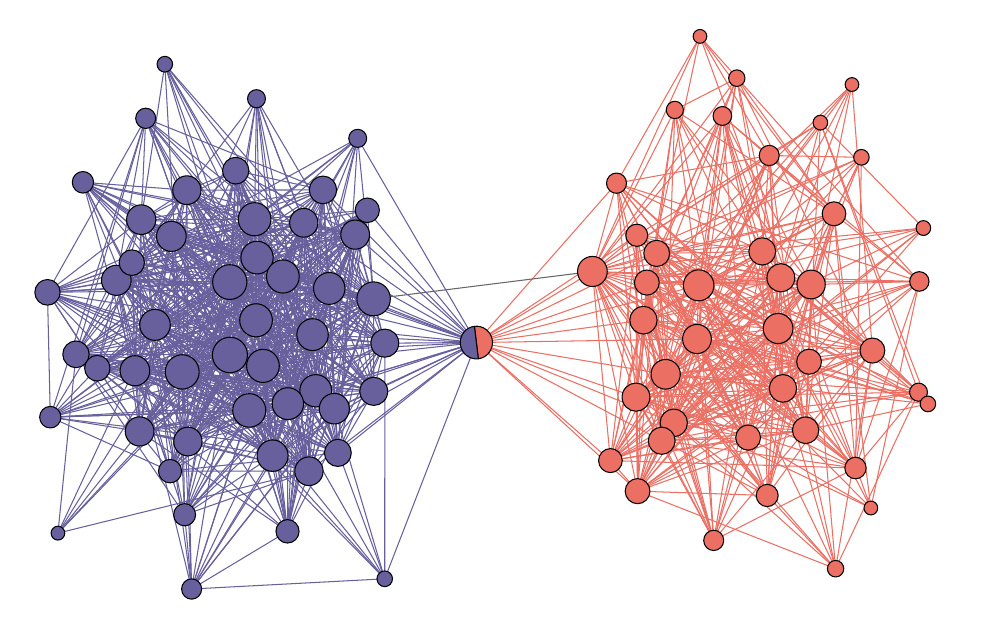}
   \caption{A real example of a pair of
   overlapped communities detected by PCMA. The overlap is
   a single vertex colored half purple and half red.
   The outbound edges from the left (right) community to
   the right (left) are all through the overlapped vertex (category $E1$),
   except for one edge that interconnects the two communities
   directly (category $E2$).
   Since an $E2$ outbound edge of a community is also an $E2$ outbound edge of the corresponding neighbor community, we adopt a simplified term that the edge is an $E2$ edge of the overlap.
     Most overlaps are like those in this example,
   having only one vertex and very few $E2$ edges, making the
   corresponding pair of communities well separated from each other.}
   \label{fig:overlap_pair}
\end{figure}

\textbf{\textit{Characteristic 4.}} How can communities ever be
distinguished when each community overlaps with a significant
number of others?  The answer is that the overlap size between two
communities is usually small, and the connection between them is
mostly through the overlap. Table~\ref{tab:dist_overlaps} lists
the frequency of occurrence of the most common overlap sizes. Out
of $232$M (millions) overlaps among the $2.9$M detected
communities, more than $80\%$ are of just a single vertex.
Fig.~\ref{fig:overlap_pair} shows the actual structure of two
detected communities.  The outbound edges from community $A$
(left) to its neighbor community $B$ are highly organized through
the overlap.  Members of $B$ usually only know the overlapping
part of $A$, and vice versa. The overlapped vertex serves as the
sole bridge and plays a unique role in passing information between
the communities.  Yet, there may exist some $E2$ edges between the
communities.  In social networks, they are possibly due to the
common member introducing members of the two communities to know
each other. In Fig.~\ref{fig:hist_c_ex}, $e_2$ is below $10\%$ or
even $5\%$ for most communities and far less than $e_1$.  It is
the small proportion of $E2$ edges that facilitates the easy
separation of communities. The proportion $e_2$ is thus an
indicator of the clearness of the boundary between a community and
its neighbor communities. We checked every pair of overlapped
communities on $E2$ edges. Results are listed in
Table~\ref{tab:dist_overlaps}. For $37.8\%$ (Sina Weibo) and
$30.1\%$ (Friendster) of them, there is not even a single $E2$
edge.  The communities maintain a good separation from their
surrounding despite each overlaps with a significant number of
neighbor communities.

\begin{table*}[htbp]
  \centering
  \caption{Distribution of overlaps as a function of size of overlaps and the
  number of $E2$ edges between a pair of communities}
  \label{tab:dist_overlaps}
  \begin{tabular*}{1.04\linewidth}{c c c c c c c c c c}
    \hline
    \multirow{2}{*}{Dataset}  & \multirow{2}{*}{No. of overlaps} & \multicolumn{4}{c}{Overlap size (no. of vertices)} & \multicolumn{4}{c}{No. of $E2$ edges of an overlap} \\ \cline{3-10}
    & & $1$  &  $2$ & $3$ & $4$ & $0$  &  $1$ & $2$ & $\leqslant 5$  \\
    \hline
    Sina Weibo & $77$ million &  $84.5\%$ &  $8.3\%$  &  $2.6\%$ &  $1.3\%$ & $37.8\%$ &  $10.5\%$  &  $6.0\%$  &  $64.0\%$ \\
    Friendster & $155$ million & $86.1\%$ &  $7.7\%$  &  $2.4\%$ &  $1.1\%$ & $30.1\%$ &  $11.0\%$  &  $6.7\%$ &  $59.4\%$ \\
    \hline
    \end{tabular*}
\end{table*}

\section*{Mesoscopic view of social network structure}

For the $2.9$ million detected communities, we can classify all
the edges in the two social networks into $5$ types (see the
caption of Fig.~\ref{fig:edge_classification}). The results are
given in Table~\ref{tab:edge_classification}. The number of Type
$1$ edges suggests that the communities account for $30\sim35\%$
of the entire network in terms of edges. These communities,
connected together by the huge number of overlaps, form an
extremely dense and tight network by themselves.  There are
$10\sim20\%$ of the edges further connecting the overlapped
communities (Type $2$).  The total number of them is comparable to
that of Type $1$, but since they are distributed among the huge
number of overlaps, each overlap shares only a very few such
edges. For example, in Sina Weibo, the 117M Type $2$ edges are
distributed among 73M overlaps, on average only $1.6$ per overlap.
The numbers further confirm the structure shown in
Fig.~\ref{fig:overlap_pair}.
The Types $1$ and $2$ edges, together occupying half of the entire
network, form an immense network of communities that can be
regarded as a hidden skeleton of the social network in the
mesoscopic scale.  The remaining half of the edges are outside the
skeleton, mostly Type $3$ or Type $4$.  The former are
``long-range'' weak ties connecting different parts of the
skeleton, thus making the skeleton an even smaller world. Although
the majority of the vertices are outside the skeleton, i.e.,
vertices with $m=0$, the edges among them (Type $5$) account for
less than $10\%$.  For Friendster it is only $1.5\%$. These
vertices are possibly the inactive users in the two online social
network services.

The edge classification helps decompose the entire network and
reveals a remarkably high proportion of the significantly
overlapped communities. The proportion could be even higher if
less tightly connected vertices are also accepted as communities.
The immense size of the network of communities confirms its
important role in social networks and invites in-depth analyses on
the properties of the huge and dense skeleton of social networks.

\begin{table}
  \centering
  \caption{Classification of edges and vertices in social networks. M represents a million. The types of edges are defined as in Fig.~\ref{fig:edge_classification}.}
  \label{tab:edge_classification}
  \begin{tabular*}{0.6\linewidth}{@{\hspace{\tabcolsep}\extracolsep{\fill}}  l | r r | r r}
    \hline
    Dataset  &  \multicolumn{2}{c|}{Sina Weibo}  &  \multicolumn{2}{c}{Friendster}  \\
    \hline
    Vertices ($m>0$)  &  $21.0$M & $26.5\%$  &  $28.0$M & $42.7\%$  \\
    Vertices ($m=0$)  &  $58.3$M & $73.5\%$  &  $37.6$M & $57.3\%$  \\
    \hline
    Edge Type 1 &  $363$M & $34.7\%$  &  $531$M & $29.4\%$  \\
    Edge Type 2 &  $117$M & $11.2\%$  &  $333$M & $18.4\%$  \\
    Edge Type 3 &  $204$M & $19.5\%$  &  $644$M & $35.7\%$  \\
    Edge Type 4 &  $273$M & $26.1\%$  &  $271$M & $15.0\%$  \\
    Edge Type 5 &  $91$M & $8.7\%$  &  $28$M & $1.5\%$  \\
    \hline
    \end{tabular*}
\end{table}

\section*{Rethinking the concept of overlapping community}

The strong empirical evidence from the analyses of the two social
networks contradicts what we usually think a community is and asks
for a reconsideration of the concept of community.  Despite a wide
variety of definitions, most of them, if not all, share an
intuitive idea: members of a community should have some sort of
internal cohesion and good separation from the rest of the
network.  The problem is how the idea should be interpreted,
especially what a good separation and the boundary of a community
are about.

Many definitions and quality measures of a community interpret
``good separation'' as the less the $k_{\mbox{\tiny
C}}^{\mathrm{out}}$ (or $k_{\mbox{\tiny
C}}^{\mathrm{out}}$/$k_{\mbox{\tiny C}}^{\mathrm{int}}$), the more
definite is the community.  Examples include the widely used
\textit{weak community} $k_{\mbox{\tiny C}}^{\mathrm{int}} >
k_{\mbox{\tiny C}}^{\mathrm{out}}$~\citep{Radicchi:2004br},
fitness function $k_{\mbox{\tiny
C}}^{\mathrm{int}}/\left(k_{\mbox{\tiny C}}^{\mathrm{int}} +
k_{\mbox{\tiny
C}}^{\mathrm{out}}\right)^{\alpha}$~\citep{Lancichinetti:2009dy,
Goldberg:2010ij}, \textit{conductance} $k_{\mbox{\tiny
C}}^{\mathrm{out}}/\left(k_{\mbox{\tiny C}}^{\mathrm{int}} +
k_{\mbox{\tiny C}}^{\mathrm{out}}\right)$ and \textit{network
community profile}~\citep{Leskovec:2009fy, Jeub:2015hs},
dynamic-based definitions such as random
walk~\citep{Rosvall:2008fi} and label
propagation~\citep{Raghavan:2007by}.  We argue that comparing
$k_{\mbox{\tiny C}}^{\mathrm{out}}/k_{\mbox{\tiny
C}}^{\mathrm{int}}$ is ineffective in large-scale networks, no
matter for overlapping or disjoint communities.  As shown in
Figs.~\ref{fig:ceio} and~\ref{fig:hist_c_ex}, there are more
outbound edges than internal edges, even if we ignore the neighbor
community edges produced by the multi-membership vertices.  The
point is that simply a larger value of $k_{\mbox{\tiny
C}}^{\mathrm{out}}$ does not necessarily mean the community is
less definite. Consider the case that an arbitrary large number of
outbound edges of a community are randomly distributed in the
whole network, the community is not really strongly connected to
any part of the network as long as the network size $n \gg
k_{\mbox{\tiny C}}^{\mathrm{out}}$.  This point has also been
discussed in a recent review by~\citet{Fortunato:2016tx}.  They
suggested using edge probabilities instead of the number of edges.
A member of a community should have a higher probability
$p_\mathrm{in}$ to form edges with the other members
than $p_\mathrm{out}$ with vertices outside the
community. Recent studies on detectability
transitions in the Stochastic Block Model~\citep{Decelle:2011ju,
Nadakuditi:2012kx, Radicchi:2013jx, Radicchi:2014kw,
Radicchi:2018kk} found that $p_\mathrm{in} > p_\mathrm{out}$ is
insufficient to guarantee that the community is detectable. There
exists a region $0 < p_\mathrm{in} - p_\mathrm{out} < \Delta$
that, although the community structure exists, no algorithms are
able to detect. It is generally difficult to infer the edge
probability between each pair of vertices.  A simplified way is to
assume the edge probabilities within a community (to the outside)
are the same and equal to the internal (outbound) edge density
$\delta_{\mbox{\tiny C}}^{\mathrm{int}} = k_{\mbox{\tiny
C}}^{\mathrm{int}}/\left[n_{\mbox{\tiny C}}(n_{\mbox{\tiny
C}}-1)\right]$, ($\delta_{\mbox{\tiny C}}^{\mathrm{out}} =
k_{\mbox{\tiny C}}^{\mathrm{out}}/\left[n_{\mbox{\tiny
C}}(n-n_{\mbox{\tiny C}})\right]$,) where $n$ and $n_{\mbox{\tiny
C}}$ are the network and community sizes, respectively.  However
for large networks $n \gg n_{\mbox{\tiny C}}$, usually
$\delta_{\mbox{\tiny C}}^{\mathrm{out}} \to 0$, making the
definition $\delta_{\mbox{\tiny C}}^{\mathrm{int}} >
\delta_{\mbox{\tiny C}}^{\mathrm{out}}$ not useful.

The problem of $k_{\mbox{\tiny C}}^{\mathrm{out}}$ (and so of
$\delta_{\mbox{\tiny C}}^{\mathrm{out}}$) is that it counts the outbound edges to
the whole network and reports only a summed quantity. What really
matters is not the number $k_{\mbox{\tiny C}}^{\mathrm{out}}$, but where the
$k_{\mbox{\tiny C}}^{\mathrm{out}}$ outbound edges are distributed. As discussed
under Characteristic 4 of the overlapping pattern, a
multi-membership vertex may contribute much to
$k_{\mbox{\tiny C}}^{\mathrm{out}}$ without messing up the boundary between the
community and its neighbors.  On the contrary, adding a number of
outbound edges to a particular vertex outside is sufficient to
change the boundary of the community. These two cases are due to
the different distribution patterns of outbound edges:
\begin{itemize}
\item Outbound edges from the same member to vertices outside the community
\item Outbound edges from different members to a particular vertex
outside the community
\end{itemize}
For the first case, it does not matter how many outbound edges
there are.  For the second case, however, the fewer the better. A
good definition of overlapping community should be able to
distinguish between the two cases. A useful concept here, as
discussed in Ref.~\citep{Xu:2017gz}, is the $f$-core -- a maximal
connected subgraph in which each vertex is connected to equal to
or more than a fraction $f$ of the other vertices in the subgraph:
\begin{equation}
b_{v,{\mbox{\tiny C}}} \geqslant f, \quad \forall v \in C
\label{eq:fcore}
\end{equation}
with $b_{v,{\mbox{\tiny C}}}$ being the belongingness of $v$ to $C$ as defined in
Eq.~(\ref{eq:belongingness}).  A vertex is acknowledged as a
member of an $f$-core as long as the vertex has sufficient
connections to the other members of the $f$-core.  It is
irrelevant whether it is connected to a large number of vertices
outside the $f$-core.  This property of $f$-core distinguishes the
two cases of outbound edges successfully and allows a vertex to
belong to multiple $f$-cores naturally.  In contrast, the
number-based counterpart called $k$-core, which requires each
vertex to be a neighbor to at least $k$ other vertices in the
subgraph, is non-overlapping by definition.  The ``maximal
connected subgraph" in the definition ensures all vertices outside
the $f$-core having belongingness less than $f$, as defined in
Eq.~\eqref{eq:fcore}, except for the case that there does exist
one vertex outside, but including it will result in some other
member(s) of the $f$-core to be kicked out. The fraction $f$
defines the boundary of the community.  A problem is that there is
no standard way to determine what value of $f$ should be used.
Communities in social networks often show core-periphery
structures~\citep{Csermely:2013bb, Rombach:2014hy, Zhang:2015eo}
and have no definite boundaries.  A large value of $f$ extracts
the core members of communities, and a small value results in more
peripheral vertices being accepted as members.  We are of the
opinion that the belongingness $b_{v,{\mbox{\tiny C}}}$ is a better way to
describe vertex memberships instead of forcing a vertex to be
either inside or outside of a community.

While the $f$-core is a good candidate, better definitions of
overlapping community may still be possible. The key point is that
the definition should take into account of the possibility of
ubiquitous presence of multi-membership vertices:
\begin{itemize}
\item The proportion of multi-membership vertices may range from
$0\sim100\%$, \item A vertex may belong to an arbitrary number of
communities,
\end{itemize}
as revealed by data analysis.  These are the causes of the
significant overlaps among communities and a much greater number
of outbound edges than internal edges.

\section*{Summary and outlook}

We studied the overlapping structure of $2.9$ million communities
detected by PCMA in the two huge online social
networks.  We found four main characteristics:
\begin{itemize}
\item Most members of a community have multiple memberships. They
are everywhere, at the periphery or in the core. \item A community
usually overlaps with a significant number of other communities,
the number typically is several times its size. \item The number
of outbound edges of a community is many times greater than the
number of internal edges. \item Although communities overlap
significantly, they remain relatively in good separation from each
other. Most overlaps concern just one or sometimes two vertices.
\end{itemize}
Note that PCMA does not impose any constraint or
implications on the fraction of overlapping vertices in a
community or the number of communities a vertex may have. It is
also capable of detecting non-overlapping or slightly overlapping
communities, as verified in the LFR benchmarking
test~\citep{Lancichinetti:2008ge} in Ref.~\citep{Xu:2017gz}.
The significant overlapping
pattern found in the two empirical social networks
asks for a rethinking of what the boundary of a community really
is.  We discussed several traditional interpretations and related
issues, and suggested the $f$-core as a possible definition for
overlapping community. Our study also showed a dense and tight
network of communities, with the communities taking the role of
vertices and the overlaps being the edges.  Most overlaps are just
of a single vertex. Each of these vertices plays a unique role in
passing on information between the communities that it belongs to.
This network of communities accounts for almost half of the entire
network.  It serves more studies on how its structural properties
would couple to many phenomena in social dynamics.

As implied by the no-free lunch theorem for
community detection that there can be no algorithm which is
optimal for all possible community detection
tasks~\citep{Peel:2017gp}, methods based on different approaches
may reveal different aspects of the community structure. In fact,
there is no standard answer to what a community really is, and it
is largely unnecessary to enforce only one definition.  This is
especially the case for empirical networks. The important thing is
that the communities detected satisfy the general notion of a
community that they have internal cohesion and relatively clear
boundaries. We verified that the $2.9$ million communities
analyzed in the present work have good separation among each
other, and high values of the intra-community edge density, as
shown in the Appendix.

In conclusion, our empirical study unfolded new aspects of
overlapping community. The results provided researchers with clues
for designing effective detection algorithms, generative models,
and benchmarks for overlapping communities, especially in social
networks. We look forward to more empirical studies powered by new
tools, to cross-check the present work and explore areas not
covered by PCMA.

\section*{Appendix: Datasets}
For completeness, we describe the two social networks we analyzed.
Table~\ref{tab:datasets} gives the basic information. Sina Weibo
is a directed network akin to Twitter.  We focused on the embedded
friendship network in which two connected individuals are
following each other. Instead of sampling small subnetworks, we
collected almost the whole giant component of the network, because
the structural completeness of the sampled network is vital to the
preservation of community structure, especially the overlapping
pattern among communities. The network data of Friendster was
downloaded from SNAP Datasets~\citep{snapnets}.

\begin{figure}[htbp]
   \centering
   \includegraphics[width=0.8\linewidth]{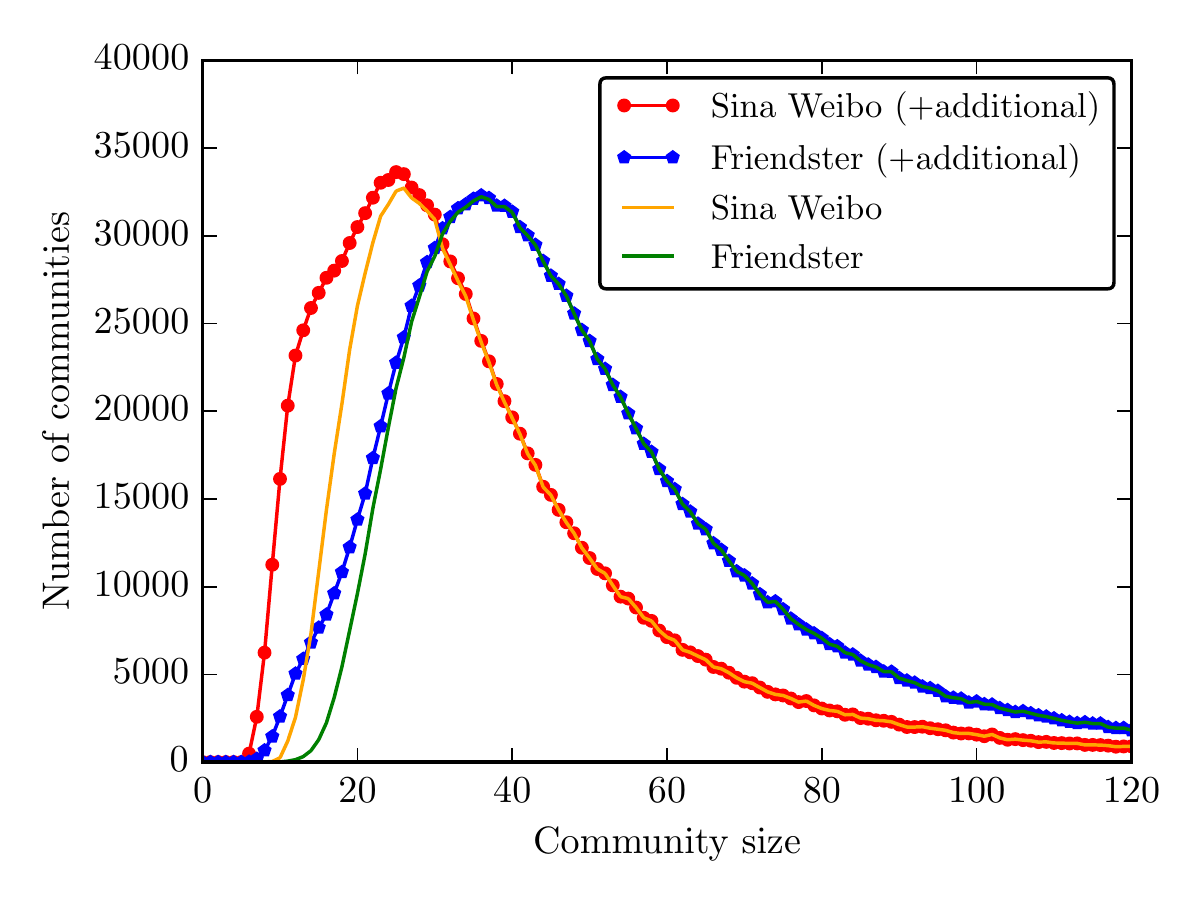}
   \caption{Size distributions of the communities detected in the two
   social networks.}
   \label{fig:c_size_dist}
\end{figure}

\begin{figure}[htbp]
   \centering
   \includegraphics[width=0.8\linewidth]{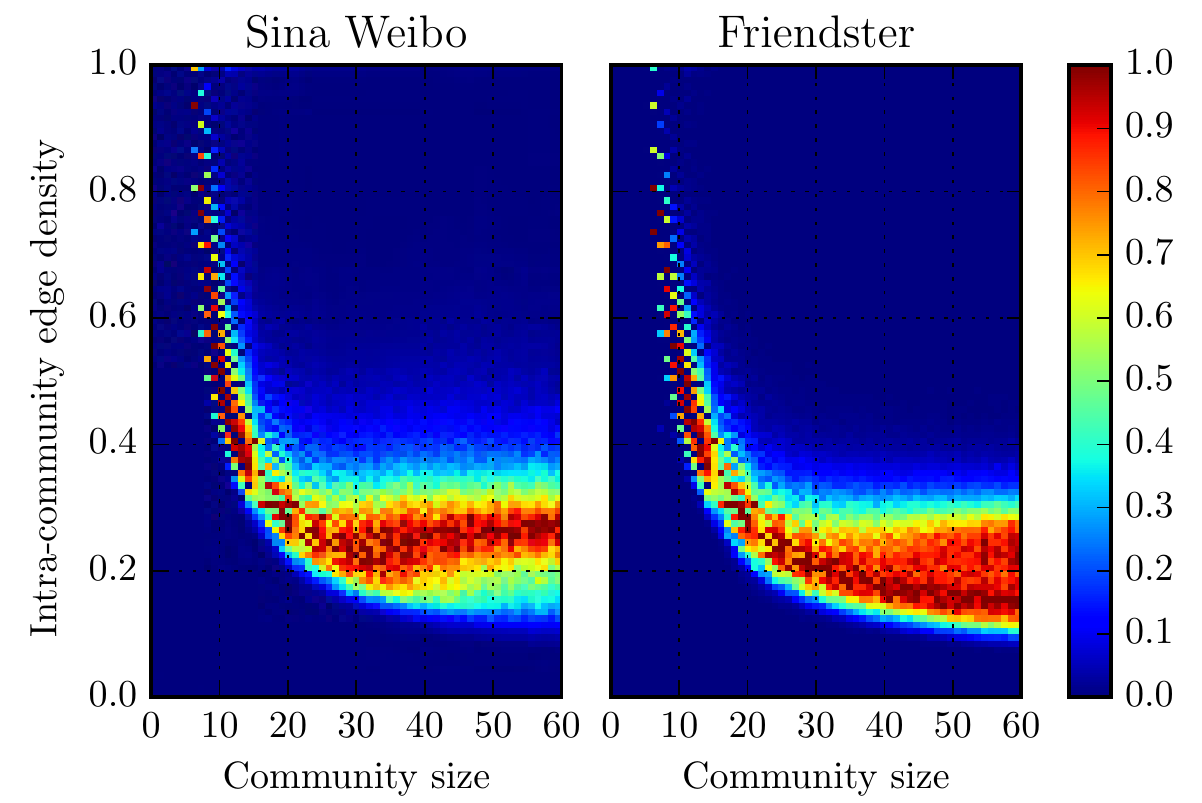}
   \caption{Histograms of the detected communities
   grouped by their size and intra-community edge density.
   The value, as given by the color, in each vertical cut are
   rescaled by mapping the highest value to unity.}
   \label{fig:c_delta_hist}
\end{figure}

We detected about $1.3$ and $1.6$ million communities in the two
networks with PCMA~\citep{Xu:2017gz}.  The algorithm is especially
suitable for detecting communities in which the vertices have
multiple memberships. Detailed information on the detection was
reported in Ref.~\citep{Xu:2017gz}. Specifically, the
three steps of PCMA were discussed in Section $2$ and the choice
of the parameters for detecting the communities within PCMA were
discussed in Appendix B of the paper. Using the symbols
introduced in Subsection 2.2 in
Ref.~\citep{Xu:2017gz}, we used a harsh threshold $l \geqslant 10$
to ensure that the detected communities are reliable (the larger
the $l$, the more reliable the community). A drawback is that many
small size communities were not included. In the present work, we
add additional communities of which $6 \leqslant l \leqslant 9$
and $g
> 3.0/l$. The results are shown in Fig.~\ref{fig:c_size_dist}.
The latter condition ensures relatively high intra-community edge
density of these communities, especially for those with low $l$.
Figure~\ref{fig:c_delta_hist} shows that all communities,
including the newly added ones, have high values of
intra-community edge density.

The large values of the proportion of intra-community and $E2$
edges, as shown in Table~\ref{tab:edge_classification}, indicate
that the number of communities we detected is close to the
possible total number of communities in the two networks. However,
it should be noted that there is no standard answer as to how many
communities there are in a real network.
We found that adding or removing the additional communities in the
analyses only produces minor changes to the statistics.  In
particular, it does not change the characteristics of the
overlapping pattern we discussed. The $2.9$ million detected
communities are believed to be adequate and representative.


\begin{backmatter}

\section*{Abbreviations}
PCMA: Partial community merger algorithm

\section*{Acknowledgements}
One of us (EHWX) gratefully acknowledges the
support from the Research Grants Council of Hong Kong SAR
Government through a Hong Kong PhD Fellowship.

\section*{Authors' contributions}
EHWX designed the research and performed the analyses. EHWX
and PMH wrote the manuscript. Both authors read and approved
the final manuscript.

\section*{Availability of data and materials}
The network data of Friendster is available in the SNAP Datasets
repository, http://snap.stanford.edu/data.

The network data of Sina Weibo was collected from weibo.com.
Due to the current terms and conditions of the platform,
the authors cannot distribute the collected data.

The source code of PCMA is available at
https://github.com/hwxu/pcma.

\section*{Funding}
The authors declare that no funding was received for the research reported.

\section*{Competing interests}
The authors declare that they have no competing interests.


\bibliographystyle{spbasic} 
\bibliography{c2}      

\begin{thebibliography}{42}
\providecommand{\natexlab}[1]{#1}
\providecommand{\url}[1]{{#1}}
\providecommand{\urlprefix}{URL }
\expandafter\ifx\csname urlstyle\endcsname\relax
  \providecommand{\doi}[1]{DOI~\discretionary{}{}{}#1}\else
  \providecommand{\doi}{DOI~\discretionary{}{}{}\begingroup
  \urlstyle{rm}\Url}\fi
\providecommand{\eprint}[2][]{\url{#2}}

\bibitem[{Backstrom et~al(2011)Backstrom, Boldi, Rosa, Ugander, and
  Vigna}]{Backstrom:2011vq}
Backstrom L, Boldi P, Rosa M, Ugander J, Vigna S (2011) {Four Degrees of
  Separation}. arXiv \eprint{1111.4570v3}

\bibitem[{Baumes et~al(2005)Baumes, Goldberg, Krishnamoorthy, Magdon-Ismail,
  and Preston}]{Baumes:2005tf}
Baumes J, Goldberg M, Krishnamoorthy M, Magdon-Ismail M, Preston N (2005)
  {Finding communities by clustering a graph into overlapping subgraphs}. In:
  Proceedings of the IADIS International Conference on Applied Computing, pp
  97--104

\bibitem[{Coscia et~al(2012)Coscia, Rossetti, Giannotti, and
  Pedreschi}]{Coscia:2012ip}
Coscia M, Rossetti G, Giannotti F, Pedreschi D (2012) {DEMON: a Local-First
  Discovery Method for Overlapping Communities}. In: Proceedings of the 18th
  ACM SIGKDD International Conference on Knowledge Discovery and Data Mining,
  ACM Press, New York, New York, USA, pp 615--623,
  \doi{10.1145/2339530.2339630}

\bibitem[{Csermely et~al(2013)Csermely, London, Wu, and Uzzi}]{Csermely:2013bb}
Csermely P, London A, Wu LY, Uzzi B (2013) {Structure and dynamics of
  core/periphery networks}. J Complex Netw 1(2):93--123,
  \doi{10.1093/comnet/cnt016}

\bibitem[{Decelle et~al(2011)Decelle, Krzakala, Moore, and
  Zdeborov{\'a}}]{Decelle:2011ju}
Decelle A, Krzakala F, Moore C, Zdeborov{\'a} L (2011) {Inference and Phase
  Transitions in the Detection of Modules in Sparse Networks}. Physical Review
  Letters 107(6):065,701, \doi{10.1103/PhysRevLett.107.065701}

\bibitem[{Epasto et~al(2017)Epasto, Lattanzi, and Paes~Leme}]{Epasto:2017jw}
Epasto A, Lattanzi S, Paes~Leme R (2017) {Ego-Splitting Framework: from
  Non-Overlapping to Overlapping Clusters}. In: Proceedings of the 23rd ACM
  SIGKDD International Conference on Knowledge Discovery and Data Mining, ACM,
  New York, NY, USA, KDD '17, pp 145--154, \doi{10.1145/3097983.3098054}

\bibitem[{Ferrara(2012)}]{Ferrara:2012ic}
Ferrara E (2012) {A large-scale community structure analysis in Facebook}. EPJ
  Data Sci 1(1):9, \doi{10.1140/epjds9}

\bibitem[{Fortunato(2010)}]{Fortunato:2010iw}
Fortunato S (2010) {Community detection in graphs}. Phys Rep 486(3-5):75--174,
  \doi{10.1016/j.physrep.2009.11.002}

\bibitem[{Fortunato and Hric(2016)}]{Fortunato:2016tx}
Fortunato S, Hric D (2016) {Community detection in networks: A user guide}.
  Phys Rep 659:1--44, \doi{10.1016/j.physrep.2016.09.002}

\bibitem[{Goldberg et~al(2010)Goldberg, Kelley, Magdon-Ismail, Mertsalov, and
  Wallace}]{Goldberg:2010ij}
Goldberg M, Kelley S, Magdon-Ismail M, Mertsalov K, Wallace A (2010) {Finding
  Overlapping Communities in Social Networks}. In: 2010 IEEE Second
  International Conference on Social Computing (SocialCom), IEEE, pp 104--113,
  \doi{10.1109/SocialCom.2010.24}

\bibitem[{Hric et~al(2014)Hric, Darst, and Fortunato}]{Hric:2014jt}
Hric D, Darst RK, Fortunato S (2014) {Community detection in networks:
  Structural communities versus ground truth}. Phys Rev E 90(6):062,805,
  \doi{10.1103/PhysRevE.90.062805}

\bibitem[{Jebabli et~al(2015)Jebabli, Cherifi, Cherifi, and
  Hamouda}]{Jebabli:2015cq}
Jebabli M, Cherifi H, Cherifi C, Hamouda A (2015) {User and group networks on
  YouTube: A comparative analysis}. In: 2015 IEEE/ACS 12th International
  Conference of Computer Systems and Applications (AICCSA), IEEE, pp 1--8,
  \doi{10.1109/AICCSA.2015.7507126}

\bibitem[{Jebabli et~al(2018)Jebabli, Cherifi, Cherifi, and
  Hamouda}]{Jebabli:2018kv}
Jebabli M, Cherifi H, Cherifi C, Hamouda A (2018) {Community detection
  algorithm evaluation with ground-truth data}. Physica A 492:651--706,
  \doi{10.1016/j.physa.2017.10.018}

\bibitem[{Jeub et~al(2015)Jeub, Balachandran, Porter, Mucha, and
  Mahoney}]{Jeub:2015hs}
Jeub LGS, Balachandran P, Porter MA, Mucha PJ, Mahoney MW (2015) {Think
  locally, act locally: Detection of small, medium-sized, and large communities
  in large networks}. Phys Rev E 91(1):012,821,
  \doi{10.1103/PhysRevE.91.012821}

\bibitem[{Lancichinetti et~al(2008)Lancichinetti, Fortunato, and
  Radicchi}]{Lancichinetti:2008ge}
Lancichinetti A, Fortunato S, Radicchi F (2008) {Benchmark graphs for testing
  community detection algorithms}. Phys Rev E 78(4):046,110,
  \doi{10.1103/PhysRevE.78.046110}

\bibitem[{Lancichinetti et~al(2009)Lancichinetti, Fortunato, and
  Kert{\'e}sz}]{Lancichinetti:2009dy}
Lancichinetti A, Fortunato S, Kert{\'e}sz J (2009) {Detecting the overlapping
  and hierarchical community structure in complex networks}. New J Phys
  11(3):033,015, \doi{10.1088/1367-2630/11/3/033015}

\bibitem[{Lancichinetti et~al(2011)Lancichinetti, Radicchi, Ramasco, and
  Fortunato}]{Lancichinetti:2011gn}
Lancichinetti A, Radicchi F, Ramasco JJ, Fortunato S (2011) {Finding
  Statistically Significant Communities in Networks}. PLoS ONE 6(4):e18,961,
  \doi{10.1371/journal.pone.0018961}

\bibitem[{Leskovec and Krevl(2014)}]{snapnets}
Leskovec J, Krevl A (2014) {SNAP Datasets}: {Stanford} large network dataset
  collection. http://snap.stanford.edu/data

\bibitem[{Leskovec et~al(2009)Leskovec, Lang, Dasgupta, and
  Mahoney}]{Leskovec:2009fy}
Leskovec J, Lang KJ, Dasgupta A, Mahoney MW (2009) {Community Structure in
  Large Networks: Natural Cluster Sizes and the Absence of Large Well-Defined
  Clusters}. Internet Math 6(1):29--123, \doi{10.1080/15427951.2009.10129177}

\bibitem[{Luccio and Sami(1969)}]{Luccio:1969hu}
Luccio F, Sami M (1969) {On the Decomposition of Networks in Minimally
  Interconnected Subnetworks}. IEEE Trans Circuit Theory 16(2):184--188,
  \doi{10.1109/TCT.1969.1082924}

\bibitem[{Lyu et~al(2016)Lyu, Bing, Zhang, and Zhang}]{Lyu:2016ia}
Lyu T, Bing L, Zhang Z, Zhang Y (2016) {Efficient and Scalable Detection of
  Overlapping Communities in Big Networks}. In: 2016 IEEE 16th International
  Conference on Data Mining (ICDM), IEEE, pp 1071--1076,
  \doi{10.1109/ICDM.2016.0138}

\bibitem[{Maiya and Berger-Wolf(2010)}]{Maiya:2010fx}
Maiya AS, Berger-Wolf TY (2010) {Sampling community structure}. In: Proceedings
  of the 19th International Conference on World Wide Web, ACM, New York, NY,
  USA, WWW '10, pp 701--710, \doi{10.1145/1772690.1772762}

\bibitem[{Nadakuditi and Newman(2012)}]{Nadakuditi:2012kx}
Nadakuditi RR, Newman MEJ (2012) {Graph Spectra and the Detectability of
  Community Structure in Networks}. Physical Review Letters 108(18):188,701,
  \doi{10.1103/PhysRevLett.108.188701}

\bibitem[{Peel et~al(2017)Peel, Larremore, and Clauset}]{Peel:2017gp}
Peel L, Larremore DB, Clauset A (2017) {The ground truth about metadata and
  community detection in networks}. Sci Adv 3(5):e1602,548,
  \doi{10.1126/sciadv.1602548}

\bibitem[{Radicchi(2013)}]{Radicchi:2013jx}
Radicchi F (2013) {Detectability of communities in heterogeneous networks}.
  Physical Review E 88(1):010,801, \doi{10.1103/PhysRevE.88.010801}

\bibitem[{Radicchi(2014)}]{Radicchi:2014kw}
Radicchi F (2014) {A paradox in community detection}. Europhysics Letters
  106(3):38,001, \doi{10.1209/0295-5075/106/38001}

\bibitem[{Radicchi(2018)}]{Radicchi:2018kk}
Radicchi F (2018) {Decoding communities in networks}. Physical Review E
  97(2):022,316, \doi{10.1103/PhysRevE.97.022316}

\bibitem[{Radicchi et~al(2004)Radicchi, Castellano, Cecconi, Loreto, and
  Parisi}]{Radicchi:2004br}
Radicchi F, Castellano C, Cecconi F, Loreto V, Parisi D (2004) {Defining and
  identifying communities in networks}. Proc Natl Acad Sci USA
  101(9):2658--2663, \doi{10.1073/pnas.0400054101}

\bibitem[{Raghavan et~al(2007)Raghavan, Albert, and Kumara}]{Raghavan:2007by}
Raghavan UN, Albert R, Kumara S (2007) {Near linear time algorithm to detect
  community structures in large-scale networks}. Phys Rev E 76(3):036,106,
  \doi{10.1103/PhysRevE.76.036106}

\bibitem[{Rees and Gallagher(2013)}]{Rees:2012jb}
Rees BS, Gallagher KB (2013) {EgoClustering: Overlapping Community Detection
  via Merged Friendship-Groups}. In: The Influence of Technology on Social
  Network Analysis and Mining, Springer Vienna, Vienna, pp 1--20,
  \doi{10.1007/978-3-7091-1346-2_1}

\bibitem[{Rombach et~al(2014)Rombach, Porter, Fowler, and
  Mucha}]{Rombach:2014hy}
Rombach MP, Porter MA, Fowler JH, Mucha PJ (2014) {Core-Periphery Structure in
  Networks}. SIAM J Appl Math 74(1):167--190, \doi{10.1137/120881683}

\bibitem[{Rosvall and Bergstrom(2008)}]{Rosvall:2008fi}
Rosvall M, Bergstrom CT (2008) {Maps of random walks on complex networks reveal
  community structure}. Proc Natl Acad Sci USA 105(4):1118--1123,
  \doi{10.1073/pnas.0706851105}

\bibitem[{Sun et~al(2017)Sun, Jie, Sauer, Ma, Han, Wang, and Xing}]{Sun:2017bw}
Sun H, Jie W, Sauer C, Ma S, Han G, Wang Z, Xing K (2017) {A Parallel
  Self-Organizing Community Detection Algorithm Based on Swarm Intelligence for
  Large Scale Complex Networks}. In: 2017 IEEE 41st Annual Computer Software
  and Applications Conference (COMPSAC), IEEE, pp 806--815,
  \doi{10.1109/COMPSAC.2017.31}

\bibitem[{Ugander et~al(2011)Ugander, Karrer, Backstrom, and
  Marlow}]{Ugander:2011ui}
Ugander J, Karrer B, Backstrom L, Marlow C (2011) {The Anatomy of the Facebook
  Social Graph}. arXiv \eprint{1111.4503v1}

\bibitem[{Watts and Strogatz(1998)}]{Watts:1998bj}
Watts DJ, Strogatz SH (1998) {Collective dynamics of `small-world' networks}.
  Nature 393(6684):440--442, \doi{10.1038/30918}

\bibitem[{Xie et~al(2013)Xie, Kelley, and Szymanski}]{Xie:2013ku}
Xie J, Kelley S, Szymanski BK (2013) {Overlapping community detection in
  networks: The state-of-the-art and comparative study}. ACM Comput Surv
  45(4):43:1--43:35, \doi{10.1145/2501654.2501657}

\bibitem[{Xu(2016)}]{Xu:2016ab}
Xu EHW (2016) {Partial Community Merger Algorithm}.
  https://github.com/hwxu/pcma

\bibitem[{Xu and Hui(2018)}]{Xu:2017gz}
Xu EHW, Hui PM (2018) {Efficient detection of communities with significant
  overlaps in networks: Partial community merger algorithm}. Netw Sci
  6(1):71--96, \doi{10.1017/nws.2017.32}

\bibitem[{Yang and Leskovec(2013)}]{Yang:2013ko}
Yang J, Leskovec J (2013) {Overlapping Community Detection at Scale: A
  Nonnegative Matrix Factorization Approach}. In: Proceedings of the 6th ACM
  International Conference on Web Search and Data Mining, ACM, New York, NY,
  USA, WSDM '13, pp 587--596, \doi{10.1145/2433396.2433471}

\bibitem[{Yang and Leskovec(2014)}]{Yang:2014fc}
Yang J, Leskovec J (2014) {Structure and Overlaps of Ground-Truth Communities
  in Networks}. ACM Trans Intell Syst Technol 5(2):26:1--26:35,
  \doi{10.1145/2594454}

\bibitem[{Yang and Leskovec(2015)}]{Yang:2015jw}
Yang J, Leskovec J (2015) {Defining and evaluating network communities based on
  ground-truth}. Knowl Inf Syst 42(1):181--213, \doi{10.1007/s10115-013-0693-z}

\bibitem[{Zhang et~al(2015)Zhang, Martin, and Newman}]{Zhang:2015eo}
Zhang X, Martin T, Newman MEJ (2015) {Identification of core-periphery
  structure in networks}. Phys Rev E 91(3):032,803,
  \doi{10.1103/PhysRevE.91.032803}

\end{thebibliography}



%

\end{backmatter}
\end{document}